\documentclass[twoside]{article}
\usepackage{tabularx}
\usepackage{epsf}
\usepackage{natbib}
\newcommand{\msun}{\ensuremath{\, {\rm M}_\odot}}
\newcommand{\n}{\ensuremath{\mem{n}}}

\newcommand{\p}{\ensuremath{\mem{p}}}

\newcommand{\cdr}{\ensuremath{^{13}\mem{C}}}
\newcommand{\czw}{\ensuremath{^{12}\mem{C}}}

\newcommand{\cvi}{\ensuremath{^{14}\mem{C}}}

\newcommand{\ndr}{\ensuremath{^{13}\mem{N}}}
\newcommand{\nvi}{\ensuremath{^{14}\mem{N}}}

\newcommand{\ovi}{\ensuremath{^{14}\mem{O}}}

\newcommand{\ose}{\ensuremath{^{16}\mem{O}}}

\newcommand{\nezw}{\ensuremath{^{22}\mem{Ne}}}

\newcommand{\nadr}{\ensuremath{^{23}\mem{Na}}}

\newcommand{\mgfu}{\ensuremath{^{25}\mem{Mg}}}
\newcommand{\mgse}{\ensuremath{^{26}\mem{Mg}}}

\newcommand{\zrse}{\ensuremath{^{96}\mem{Zr}}}
\newcommand{\zrfu}{\ensuremath{^{95}\mem{Zr}}}
\newcommand{\zrvi}{\ensuremath{^{94}\mem{Zr}}}

\newcommand{\mem}[1]{\ensuremath{\mathrm{ #1}}}
\newcommand{\sprn}{\mbox{$s$ process}}
\newcommand{\spr}{\mbox{$s$-process}}
\newcommand{\kelv}{\ensuremath{\,\mathrm K}}
\newcommand{\abb}[1]{Fig.\,\ref{#1}}
\newcommand{\kap}[1]{\S\,\ref{#1}}

\begin{document}

\setcounter{figure}{0}
\setcounter{section}{0}
\setcounter{equation}{0}

\begin{center}
{\Large\bf
The Second Stars}\\[0.7cm]

Falk Herwig\\[0.17cm]
Los Alamos National Laboratory \\
Los Alamos, Theoretical Astrophysics Group, NM 87544, USA  \\
fherwig@lanl.gov
\end{center}

\vspace{0.5cm}

\begin{abstract}
\noindent{\it The ejecta of the first probably very massive stars
polluted the Big Bang primordial element mix with the first heavier
elements. The resulting ultra metal-poor abundance distribution
provided the initial conditions for the second stars of a wide range
of initial masses reaching down to intermediate and low masses. The
importance of these second stars for understanding the origin of the
elements in the early universe are manifold.  While the massive first
stars have long vanished the second stars are still around and
currently observed. They are the carriers of the information about the
first stars, but they are also capable of nuclear production
themselves. For example, in order to use ultra or extremely metal-poor
stars as a probe for the r-process in the early universe a reliable
model of the s-process in the second stars is needed.  Eventually, the
second stars may provide us with important clues on questions ranging
from structure formation to how the stars actually make the elements,
not only in the early but also in the present universe. In particular
the C-rich extremely metal-poor stars, most of which show the \spr\
signature, are thought to be associated with chemical yields from the
evolved giant phase of intermediate mass stars. Models of such AGB
stars at extremely low metallicity now exist, and comparison with
observation show important discrepancies, for example with regard to
the synthesis of nitrogen. This may hint at burning and mixing aspects
of extremely metal-poor evolved stars that are not yet included in the
standard picture of evolution, as for example the hydrogen-ingestion
flash. The second stars of intermediate mass may have also played an
important role in the formation of heavy elements that form through
slow neutron capture reaction chains (s-process). Comparison of models
with observations reveal which aspects of the physics input and
assumptions need to be improved. The s-process is a particularly
useful diagnostic tool for probing the physical processes that are
responsible for the creation of elements in stars, like for example
rotation. As new observational techniques and strategies continue to
penetrate the field, for example the multi-object spectroscopy, or the
future spectroscopic surveys, the extremely metal-poor stars will play
an increasingly important role to address some of the most fundamental
and challenging, current questions of astronomy.}
\end{abstract}

\section{Introduction}
\label{sec:intro}
One of the most intriguing questions of astrophysics and astronomy is
 the origin of the elements in the early universe, and how this
 relates to the first formation of structure.  About 200,000 years
 after the Big Bang the epoch of structure formation emerged and the
 first stars were born from the initial, pristine baryonic matter.
 Without any elements heavier than helium to provide cooling, the
 first stars that formed from the baryonic matter trapped in the
 emerging mini-dark matter halos were probably very massive, greater
 than 30\msun\ \citep[e.g.][]{abel:02}.  These massive stars burned
 through their available fuel in about one to two million years,
 exploded as supernovae, and dispersed the first elements heavier than
 helium into the nascent universe, or collapsed into black
 holes. These first events of stellar evolution influenced their early
 universe neighborhood, and determined under which conditions and with
 which initial abundance low-mass stars with masses like the Sun
 eventually formed. These take about 100 to 1000 times longer to form
 than more massive stars.  Thus, by the time the first low-mass stars
 formed there was already a small amount of heavier elements
 present. Such low-mass stars are really the second stars, stars which
 formed from a non-primordial abundance distribution.  These second
 stars are important because while the massive first stars have long
 since vanished, some second stars are still around.  Their importance
 lies in their capacity as carriers of the information about the
 formation and evolution of those long gone first generations of
 stars. The second stars are also important because of the nuclear
 production they are capable of by themselves (for example the first
 s-process in the second stars).  Observations of such second stars
 are now emerging in increasing quantity and with high-resolution
 spectroscopic abundance determination \citep{beers:05}.

There are a number of strategies to find ultra and extremely
metal-poor stars. The most effective appears to be spectroscopic
surveys. Most of the EMP and UMP stars have been discovered in the HK
survey \citep{beers:92} and the Hamburg/ESO survey
\citep[HES][]{christlieb:01}.  Here, I designate somewhat arbitrarily
ultra metal-poor (UMP) stars as those with $\mem{[Fe/H]} \leq -3.5$
($Z\leq 6\cdot 10^{-6}$) and extremely metal-poor (EMP) stars as those
with $\mem{[Fe/H]} \leq -1.8$ ($Z\leq 3\cdot 10^{-4}$). EMP stars
include the most metal-poor globular clusters. UMP stars are rare and
hard to find in the spectroscopic surveys. Only approximately a dozen
have been reported so far, with the present record holder HE 0107-5240
at $\mem{[Fe/H]} = -5.4 $ \citep{christlieb:02,christlieb:04}. At and
below this metallicity spectral features are so weak that even the 10m
class telescopes are barely suitable for the job.

The formation mode of the second stars depends on a complex history of
events, each of which may leave signatures in the observed abundances
of EMP and in particular the UMP stars. These events include the
ionizing radiation of the very first primordial stars and the
expansion of the associated H II regions, their effect on possibly
primordial star formation in neighboring mini-dark matter halos, as
well as possibly the metal enrichment from a pair-instability
supernova, depending on the mass of the first primordial stars. These
processes and the cosmological assumption that fix the underlying
dark-matter structure will eventually determine the abundance
distribution of the EMP and UMP stars now discovered and analyzed. The
most recent studies confirm that the very first primordial stars are
 massive \citep{oshea:05}. However in these calculations the
effect of the ionization from star formation in nearby mini-dark
matter halos is taken into account, and it is found that the following
generation of primordial stars may include intermediate mass stars.

 \begin{figure}[t]
\epsfxsize=9cm
\center{ \epsfbox{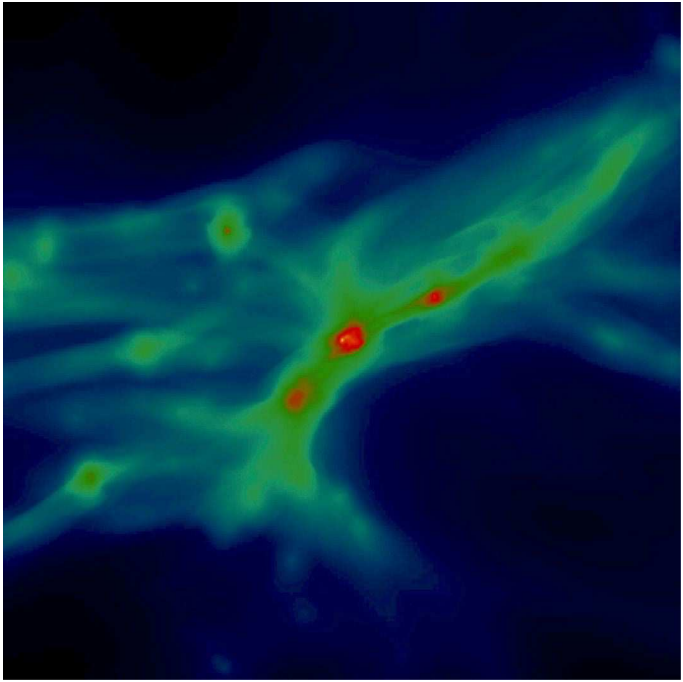}}
\caption{ \label{fig:dmhalo} Distribution of baryonic matter
  clustering around a cosmological dark matter halo in a
  hydrodynamics and N-body simulation at redshift $z \approx 17$
  \citep{oshea:05}. The projection volume of $1.5$ proper kiloparsecs
  on a side is centered on the halo where the first population III
  star in this region will form.  This halo is surrounded by other
  halos in which the second generations of primordial and/or
  metal-enriched stars will form. Star formation in these neighboring
  halos is effected by the ionizing radiation from the first star
  formation in the center halo, and may include primordial or
  ultra metal-poor intermediate mass stars.}
\end{figure}
The most metal-poor stars are potentially an extremely powerful tool
to study star formation and evolution in the early universe
\citep{beers:05}. These stars with metallicities reaching in excess of
5 orders of magnitude in metallicity below the solar metallicity often
show abundance distributions that are very different from the solar
abundance distribution. The solar abundance distribution is the result of a long
galactic chemical evolution, reflecting the cumulative result of mixing and
nuclear processing of many generations of stars of a wide range of
masses. The most metal-poor stars are very old
and have formed very early, at the onset of galactic chemical
evolution. It is therefore reasonable to assume that the abundance
pattern in these second stars are the result of only few different
sources and corresponding stellar nuclear production sites. The goal
is to identify these nuclear production sites and the mechanisms and
processes through which the corresponding stellar ejecta have been
brought together in the star formation cloud of the second stars. An
important part of the problem is to account for self-pollution or
external pollution events that may have altered the stellar abundances
between the time of second star formation and observation.

As the number of EMP and UMP stars with detailed abundance
continuously grows one can discern certain abundance patterns. A very
unusual sub-group are the carbon-rich EMP and UMP stars. Observations
indicate that many of these stars are polluted by Asymptotic Giant
Branch (AGB) binary companions. This implies that reliable models of
AGB stars including their heavy element production through the
s-process at extremely low metallicity are required to understand
these stars, and to disentangle the AGB contribution from other
sources that have contributed to the observed abundance patterns. In
the following section the carbon-rich EMP and UMP stars will be
discussed in more detail. In \kap{sec:improve} current models of EMP
intermediate mass stars and their uncertainties are presented. Such
models have important applications for the emerging field of near-field
cosmology, which uses galactic and extra-galactic stellar abundances
to reconstruct the processes that lead to the formation of our and
other galaxies (\kap{sec:nfc}). Then, ongoing work -- both
observational and theoretical -- to determine the nucleosynthesis of
nitrogen at extremely low metallicity is described
(\kap{sec:nitrogen}) and \kap{sec:hingba} covers the
hydrogen-ingestion flash in born-again giants, because this type of
events may be important for EMP AGB stars as well, and may indeed be a
major source of nitrogen in these stars. \kap{sec:spr} deals with the
\sprn, and how it can be used as a diagnostic tool to improve the
physics understanding of stellar mixing. Finally, and before some
forward lookgin concluding remarks (\kap{sec:concl}) the importance of
nuclear physics input for the question of C-star formation will be
discussed in \kap{sec:enrt}.

\section{The C-rich most metal-poor stars}
\label{sec:CEMP}
About $20 \dots 30\%$ of all EMP and UMP stars show conspicuous
enrichment of the CNO elements, most notably C. For example, CS
29497-030 \citep{sivarani:04a} has $\mem{[Fe/H]} = -2.9$,
$\mem{[C/Fe]} = 2.4$, $\mem{[N/Fe]} = 1.9$, $\mem{[O/Fe]} = 1.7$, and
$\mem{[Na/Fe]} = 0.5$. Others, like LP 625-44 \citep[$\mem{[Fe/H]} =
-2.7$][]{aoki:02} or CS 22942-019 \citep[$\mem{[Fe/H]} =
-2.7$][]{preston:01}, have a similar overabundance of C, but smaller N
($\mem{[N/Fe]} = 1.0$ for LP 625-44) and larger Na ($\mem{[Na/Fe]} =
1.8$ for LP 625-44). Many of the heavier elements, like Ti, Cr, Mn or
Zn in CS 29497-030 are either not overabundant or rather somewhat
underabundant compared to the solar distribution.

Large overabundances of N and Na are not predicted by standard models
of massive stars, neither for Pop III \citep{heger:02a} nor at larger
metallicity \citep{woosley:95}.  Primary production of N requires to
expose the He-burning product C to H-burning again. Such a condition
is encountered in massive AGB stars during hot-bottom burning
\citep{herwig:04c}. Na is made in C-burning in massive stars but its
final abundance is sensitive to the neutron excess, and thus scales
less than linear with metallicity. If primary \nvi\ is exposed to
He-burning it will lead to the production of \nezw. Either an
additional proton collected in hot-bottom burning, or a neutron from
the $\nezw(\alpha,\n)\mgfu$ reaction in the He-shell flash of AGB
stars can then lead to primary \nadr\ production. This process leads
to significant Na overabundances in standard AGB stellar models of
very low or zero metallicity \citep{herwig:04a,siess:04}.

The overabundances of C, N and Na are not the only indication that the
abundances in C-rich EMP stars are at least partly caused by AGB
stellar evolution. All of the objects mentioned above, and in fact
most C-rich EMP stars show in some cases significant overabundances of
the s-process elements. In addition they reveal their binarity through
radial velocity variations, corresponding to binary periods of a few
days in the case of HE 0024-2523 \citep{lucatello:02} to more typically
one to a dozen years. In fact, in their statistical analysis
\citet{lucatello:04a} find that observations of radial velocities of
CEMP-s stars (C-rich EMP stars with s-process signature) obtained to
date is consistent with a $100\%$ binarity rate of CEMP-s stars.

HE 0107-5420 is currently the intrinsically most metal-poor star with
$\mem{[Fe/H]} = -5.3$. However, the total metal abundance in terms of
elements heavier than He is quite large. In particular the CNO
elements are significantly overabundant compared to the solar
abundance ratios, with $\mem{[C/Fe]} = +4.0$, $\mem{[N/Fe]} = +2.4$
\citep{christlieb:04}, and $\mem{[O/Fe]} = +2.3$ \citep{bessel:04}.
The large overabundance alone indicate that only a few nuclear sources
were involved in creating this abundance pattern. Together with the
low Fe abundance the large CNO abundances require a primary production
of the CNO elements. A primary nucleosynthesis production chain in a
star is based only on the primordial elements H and He. A secondary
production in contrast requires heavier elements already to be present
in the initial element mix at the time of star formation. For example,
standard models of massive stars predict a secondary production of N
from CNO cycling of C and O that is initially present from earlier
generations of stars. 

HE 0107-5420 is a good example to discuss the possible nuclear
production sites that may be important at the lowest metallicities.
Certainly the production of the first massive Pop III stars plays an
important role. \citet{umeda:03} have discussed the observed abundance
pattern within the framework of mixing and fallback of the ejecta
during the explosion of supernovae that eventlually form a black hole.
They show that the mixing and fallback parameters can be chosen in a
way that accounts for much of the observed pattern. For example, their
model correctly reflects the large observed Fe/Ni ratio, and the large
C/Fe and C/Mg ratios. The model does predict a significant N
overabundance but it quantitatively falls short of matching the
observed value. Their model predicts the charactereistic odd-even
effect that reflects the small (or missing) neutron-excess of the
primordial nuclear fuel. One of the observed species that hints that
HE 0107-5420 may not have such a pronounced odd-even pattern is Na.
The observerved overabundance is $\mem{[Na/Fe]} = 0.8$, more than 1.5
dex above the model prediction from the mixing and fallback supernova.
Unfortunately, for the heavy trans-iron elements only upper limits
could be derived so far, which prohibits definite access to the s- and
r-process abundances in HE 0107-5420. At this point Na and to a lesser
extent N are the major indicators that some additional sources are
part of the nuclear production site inventory that caused the
abundance pattern in this most metal-poor star.

Without better guidance from ab-initio coupled structure and star
formation simulations -- such as shown in \abb{fig:dmhalo} -- several
scenarios have to be considered.  Material from the first supernova
could be ejected from the corresponding halo and injected into a
neighbouring halo. This metal enrichment could lead to the formation
of an intermediate mass star, for example a very massive AGB star,
maybe even a super-AGB star that eventually ends in a ONeMg
core-collapse supernova. As shown in \kap{sec:improve} a contribution
from such stars can account for both Na as well as N, but more
detailed models of this particular scenario need to be done for
quantitative comparison. A low-mass star like HE 0107-5420 could then
form from the combined nuclear production of these two sources, a
roughly $25\msun$ black hole forming supernova and a massive AGB or
super-AGB star.

Alternatively the low-mass star forms from the ejecta of just one
supernova, and this initial abundance pattern is modified by
self-polution.  HE 0107-5420 is in fact a giant on the first ascent.
Observations, in particular of globular cluster red giants show that
mixing and nuclear processing of C into N below the deep convective
envelope lead to secondary N production
\citep{denissenkov:03b,weiss:04}. This could explain the quantitative
discrepancy between the observed N abundance and the model by
\citet{umeda:03}. However, the rather large carbon isotopic ratio
$\czw/\cdr \sim 60$ implies that self-polution can not account for the
observed Na abundance, in particular assuming the absence of \nezw\
which is a secondary nucleosynthesis product in massive stars.

This leads to the external polution scenario that is frequently
considered in this context. As mentioned above many CEMP stars are in
fact in binary systems, and the abundance signatures seen in these
C-rich very metal-poor stars could be -- in part -- the result of mass
transfer from the AGB progenitor of a white dwarf companion. For HE
0107-5420 \citet{suda:04} discuss this possibility in detail. For the
evolution of primordial AGB stars they consider the possibility of the
H-ingestion flash, a process in which the He-shell flash convection
zone reaches outward into the H-rich envelope, resulting in a peculiar
nucleosynthesis and mixing regime. Such events are known at solar
metallicity to cause a significant fraction of young white dwarfs to
evolve for a short period of time back to the AGB (born-again
evolution, \kap{sec:hingba}), and the most prominent representative of
this class of objects is Sakurai's object \citep{duerbeck:00}.

\section{Stellar evolution models for the second stars}
\label{sec:improve}
As discussed in the previous section the evolution of low- and
intermediate mass stars at zero or extremely or ultra metal-poor
abundance is essential for studying the most metal-poor stars and
their cosmological origin and environment. This section deals first
with the evolution of AGB stars that may have formed from Big Bang
material. A possibly important evolutionary phase - the H-ingestion
flash triggered by the He-shell flash - has been observed in models of
Pop III and even in models of non-zero but ultra-low metal content
(\kap{sec:hflash}).  Finally the evolution and yields of extremely
metal-poor TP-AGB stars is described. The evolution of Pop III massive
stars is discussed, for example, by \citet{heger:02a}.

\subsection{Pop III AGB evolution}
\label{sec:z00}
Simulations show that many of the overall properties of Pop III
intermediate mass stellar models \citep{chieffi:01,siess:02} probably
apply to ultra metal-poor stars as well. At $Z=0$ the initial
thermal pulses may show peculiar convective mixing events at the
core-envelope boundary during or after the actual He-shell flash. But
eventually the models enter a phase of rather normal thermal pulse AGB
evolution, with regular thermal pulse cycles, third dredge-up,
hot-bottom burning and mass loss. Many uncertainties of the Pop III
stellar evolution is related to the possibility of flash-burning that
is not yet well understood (\kap{sec:hflash}).

$Z=0$ stellar evolution is different from evolution at higher metal
content because CNO catalytic material for H-burning is initially
absent. Burning hydrogen via the pp-chain requires a higher
temperature than H-burning via the CNO cycle. Eventually the
triple-$\alpha$ process will provide some C, and a mass fraction as
low as $10^{-10}$ is sufficient to switch H-core burning to CNO
cycling. During He-core burning some C is produced outside the
convective core and a tail of carbon reaches from the convective core
out toward the H-shell.  After the end of He-core burning, and
depending on the initial mass, this carbon and some nitrogen is mixed
into the envelope. According to \citet{chieffi:01} this second
dredge-up raises the envelope C-abundance above $10^{-7}\msun$ for
$M_{ini}>6\msun$.  For such a C-abundance the H-shell is fully
supported by CNO cycling.  As a result the thermal pulse AGB evolution
is like that of the ultra metal-poor cases.

Models with initial masses below $6\msun$ show a peculiarity that is
unknown in more metal-rich models. After an initial series of weak
thermal pulses the H-shell forms a convection zone when it re-ignites
after the He-shell flash \citep{chieffi:01,siess:02,herwig:03a}. The
lower boundary of this convection zone is highly unstable, because it
coincides with the opacity discontinuity that marks the core-envelope
interface.  Even small amounts of mixing will drive flash-like burning
and deep mixing, as protons enter a \czw-rich zone. It seems that the
details of the evolution of the H-shell convection zone is not
important as this one-time event leads to deep dredge-up and
enrichment of the envelope with a sufficient amount to support regular
CNO cylce burning in the H-shell thereafter.  Accordingly, these
lower-mass cases will evolve like extremely metal-poor stars too, and
the $Z=0$ models discussed here can be used as proxies for what is in
reality born as an ultra metal-poor star.

The effect of rotationally induced mixing processes may alter the
evolution of low- and intermediate-mass stars and very low metallicity
even before the the AGB and lead to a qualitatively different thermal
pulse evolution.  \citet{meynet:02} describe models for metallicity
$Z=10^{-5}$ in which \czw\ and \ose\ is mixed out of the He-burning
core and up to the location of the H-shell. This leads to an increase
of CNO material, in particular \nvi\ in the envelope that could result
in a rather normal thermal pulse evolution without any peculiar
hydrogen convection zone. Whether the rotationally induced mixing
processes before the AGB or the subsequent thermal pulse AGB evolution
with the associated dredge-up events dominate the final CNO yields is
not yet clear.

The difference between a real Pop III and an extremely metal-poor
thermal pulse AGB star is the initial absence of Fe and other elements
heavier than Al in the Pop III star. This does not preclude that a Pop
III AGB stars can not generate \spr-elements by the slow neutron
capture process.  \citet{goriely:01} have used the \spr-framework
established for solar-like metallicities (\kap{sec:spr}) and ran
network calculations with an initial abundance appropriate for the
interior of a $Z=0$ TP-AGB star. They used the models of
\citet{siess:02} as stellar evolution input. The major uncertainty was
the mixing for the formation of the neutron donor species \cdr, and
the unknown feedback of such mixing to the thermodynamic structure
evolution.  \citet{herwig:03c} and \citet{goriely:04} have shown that
this feedback is likely very important. Nevertheless, the \spr-models
for $Z=0$ are instructive, because they show that if a sufficient
neutron source is available, the \sprn\ can be based on seed nuclei
with lower mass number than Fe, for example C. C is produced in a
primary mode in TP-AGB stars.  For such a $Z=0$ \sprn\ the Fe/Ni ratio
should be markedly sub-solar, approximately close to the quasi
steady-state value of $\sim3$, that is given by the ratio of the
neutron cross sections of Ni and Fe (the isotopic minimum for each
element). HE 0107-5240 with $\mem{[Fe/H]}=-5.3$ \citep{christlieb:04}
shows a larger than solar Fe/Ni ratio, which is evidence for the
contribution of supernovae to the abundace distribution of this star.

\subsection{The hydrogen-ingestion flash in ultra metal-poor or metal-free AGB stars }
\label{sec:hflash}
The peculiar H-convective episode described in the previous section
was observed in AGB models with of $5\msun$ and $Z=0$ by
\citet{herwig:03a}. During the following thermal pulse the He-shell
flash convection zone reachea out into the H-rich envelope. This leads
to the H-ingestion flash (HIF). Protons are mixed on the convective
time scale into the He-shell flash region which is with depth
increasingly hot.  The hydrogen flash-burning leads to a separate
convection zone as shown in \abb{fig:ff08348}. Convective H-burning in
this layer is characterized by a large abundance of \czw\ and protons.
The protons have been mixed into this region from the envelope.
Accordingly the \nvi\ abundance in this layer is very large, up to
$10\%$ by mass in a layer of $\approx 10^{-3}\msun$. The models show
that after both the He-shell flash and the HIF have subsided a deep
dredge-up episode will mix this \nvi-rich layer into the envelope.
This may account for a substantial \nvi\ production and part of the
observational pattern observed in EMP stars that are believed to be
poluted by EMP AGB stars (\kap{sec:nitrogen}).
 A HIF was also found in the $2\msun$,
$Z=0$ model, but \citet{herwig:03a} did not find the HIF in models of
$Z\geq10^{-5}$.
 \begin{figure}[t]
\epsfxsize=13cm
\center{\epsfbox{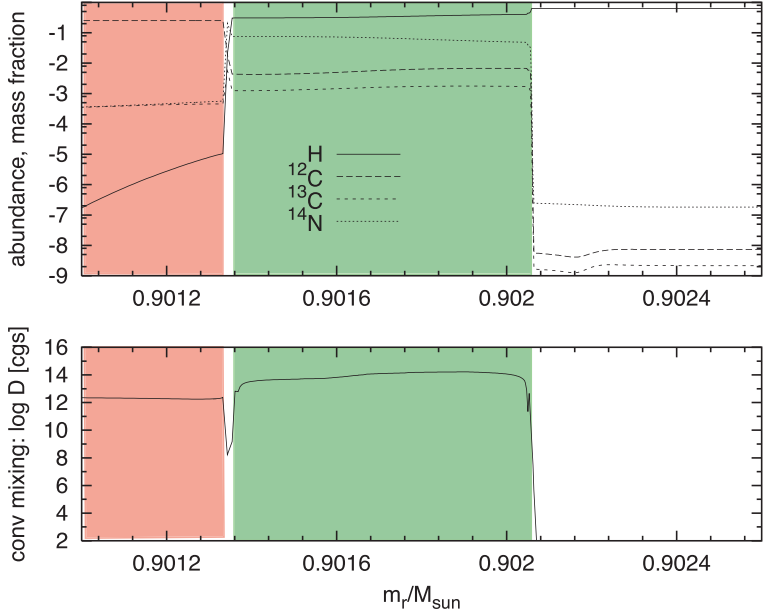}}
\caption{ \label{fig:ff08348} Burning and mixing in the H-ingestion
  flash during the 10$^{\mem{th}}$ thermal pulse of a $5\msun$, $Z=0$
  sequence \citep{herwig:03a}.  Shown are the top layers of the
  He-shell flash convection zone (red), and the H-ingestion
  flash-driven convection zone (green). Top panel: abundance profiles;
  bottom panel: convective mixing coefficient.  }
\end{figure}

AGB stars are highly non-linear systems. The occurrence of the HIF
depends on many details of the models, including the physics of
mixing, opacities, the numerical resolution and nuclear reaction
rates. Small changes in any of these ingredients, that may go
unnoticed during a more robust stellar evolution phase (like the
main-sequence), can be the deciding factor here whether a HIF occurs
or not. Currently, theoretical models do not agree if and in which
initial mass and metallicity regime the HIF occurs
\citep{fujimoto:00,chieffi:01,siess:02,herwig:03a}. However, it is
important to better understand this situation because the HIF may play
an important role in the \sprn\ in extremely metal-poor stars
\citep{iwamoto:04}. Hydrogen ingested into the He-shell flash or
pulse-driven convection zone (PDCZ) may lead to \cdr\ production which
could release neutrons. \citet{suda:04} propose that the abundance
pattern of the currently most metal-poor star -- HE 0107-5420 -- is in
part due to the nucleosynthesis in a thermal pulse AGB star with HIF.
The details of the nucleosynthesis in a HIF depend sensitively on the
physics of simultaneous rapid nuclear burning and convective mixing.
The hydrogen-ingestion flash can also be a source of lithium,
distinctly different from conditions during hot-bottom burning
\citep{herwig:00f}.

The evolution of born-again stars, like Sakurai's object, provide
valuable additional constraints with regard to the HIF in AGB stars
(\kap{sec:hingba}). The very late thermal pulse is associated with a
HIF \citep{iben:83a,herwig:99c}.  By reproducing the evolution of stars
like Sakurai's object one can gain some confidence in models of the
HIF in metal-poor AGB stars.  \citet{herwig:01a} showed that models in
which the convective hydrogen ingestion occurs with lower velocities
than predicted by the mixing-length theory \citep{boehm-vitense:58}
feature a faster born-again evolution than MLT models, quantitatively
in agreement with observations.  The physical interpretation is that
rapid nuclear burning on the convective time scale releases energy and
adds buoyancy to down-flowing convective bubbles, leading to
additional breaking of the plumes. The temperature and time scales for
nucleosynthesis during the HIF would be significantly different.  This
has not yet been applied to HIF models at $Z=0$ or extremely low
metallicity.

\subsection{AGB nucleosynthesis at extremely low metallicity}
\label{sec:nucEMP}
By definition all yields of Pop III stars are primary, because the
initial composition contains only Big Bang material. Nuclear production
is secondary when nuclei heavier than H and He are already present in
the initial abundance distribution and transformed in other species.
POP III or EMP AGB stars can produce a large number of species in
primary mode, including the CNO elements, Ne, Na, Mg and Al. The
production of heavier elements depends on the availabilitity of
neutrons.  Some of these primary species, like oxygen, are usually not
considered a product of AGB evolution at moderate metal-deficiency or
solar metallicity.  However, the initial envelope mass fraction of
oxygen for models with $Z=10^{-4}$ is $<5\cdot10^{-5}$.  \ose\ in the
intershell material (cf.\ \abb{fig:kipp}) that is dredged-up is
primary and even at this low metallicity of the order $1\%$ by mass.
Dredge-up of such material will enhance the envelope abundance
significantly.
\begin{figure}[t]
\epsfxsize=12cm
\center{ \epsfbox{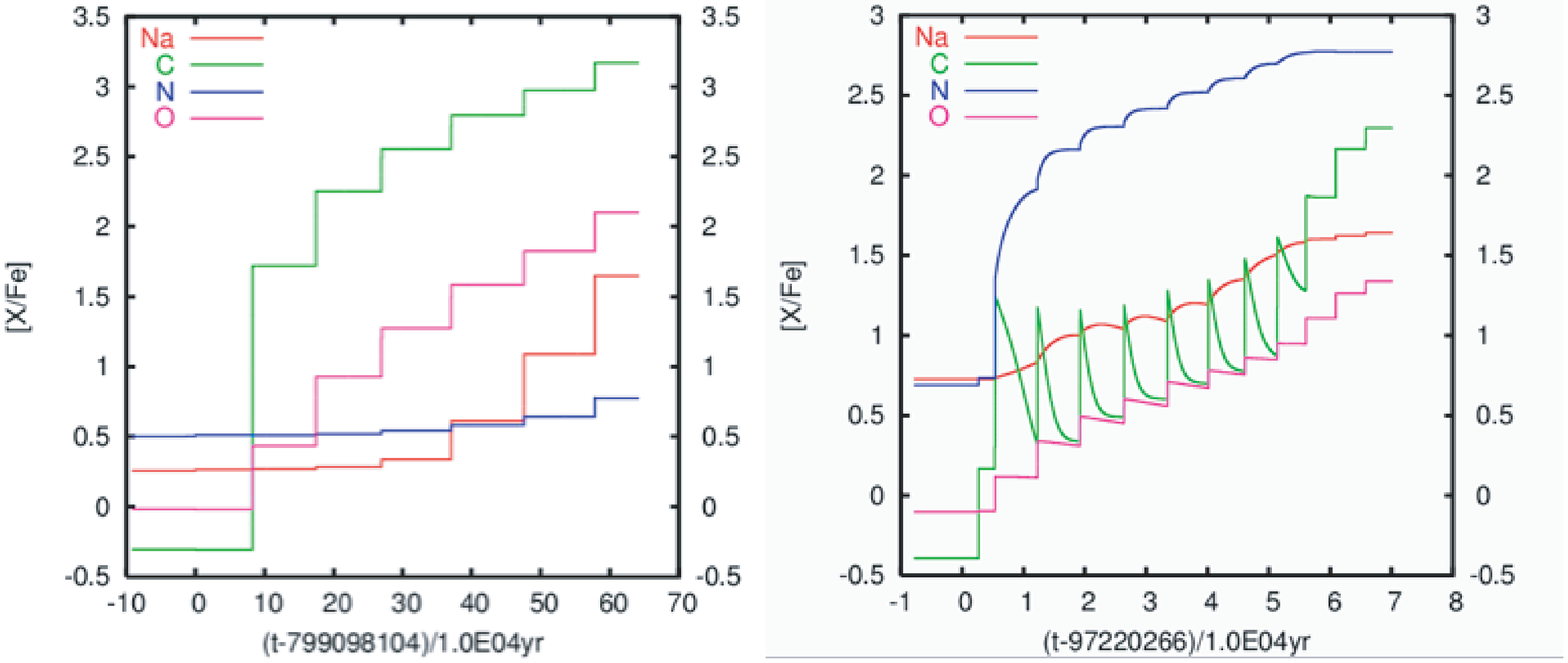}}
\caption{ \label{fig:NaCNO} 
  Surface evolution of C, N, O and Na for 2 and 5\msun\ AGB evolution
  models at $\mem{[Fe/H]}=-2.3$ \citep{herwig:04a}. The ejecta of the
  2\msun\ case are released after $0.8\mem{Gyr}$, whereas the $5\msun$
  case releases the ejecta after $0.1\mem{Gyr}$. The abundance
  evolution is determined by the interplay of dredge-up and hot-bottom
  burning. Each step in the curves corresponds to a dredge-up event
  after a thermal pulse. Hot-bottom burning can be obseved in the
  5\msun\ case as the decrease of C and increase of N (and to a lesser
  extent Na) inbetween dredge-up events. Note the dichotomy in the C/N
  ratio in the low-mass vs.\ high-mass AGB star. }
\end{figure}

A set of low- and intermediate mass AGB stellar evolution models with
$Z=10^{-4}$ ([Fe/H] = -2.3) with detailed structure, nucleosynthesis
and yield predictions has been presented by \citet{herwig:04a}. The
oxygen overabundance in these models is $\log(X/X_\odot)=0.5 \cdots
1.5$, depending on mass. The abundance evolution for C, N, O and Na is
shown in \abb{fig:NaCNO} for two initial masses. The evolution of C and
N is qualitatively different in the two cases. The lower mass sequence
shows the increase of C and some O due to the repeated dredge-up
events. No significant amount of \nvi\ is dredged-up. The result is a
large C/N ratio. The $5\msun$ track shows the opposite behaviour. The
much lower C/N ratio is the result of hot-bottom burning
\citep{boothroyd:93}. C and O are dredged-up after the thermal pulse
as in the lower mass case, but during the following quiescent
intperpulse phase the envelope convection reaches into the region hot
enough for H-burning and the envelope C is transformed into N. The
initial-mass transition between hot-bottom burning models with large N
abdundances and C dredge-up models without hot-bottom burning is very
sharp. This precludes the notion that EMP stars with simultaneously
large N and C overabundances, like CS 29497-030 with
$\mem{[C/Fe]}=2.4$ and $\mem{[N/Fe]}=1.9$ \citep{sivarani:04a}, may be
poluted by such AGB stars in the transition regime where hot-bottom is
only partially efficent (see \kap{sec:nitrogen}).

The signature of hot-bottom burning in \abb{fig:NaCNO} is a gradual
change of abundance between the steps caused by dredge-up. From this
it can be seen that Na has a different nuclear production site
depending on initial mass. In the $2\msun$ case Na is produced in the
He-shell flash, by n-captures on \nezw. The neutrons come from the
$\nezw(\alpha,\n)$ reaction. \nadr\ is the dredged-up but not produced
during the interpulse phase. The neutron-heavy Mg isotope \mgse\ is
produced in a similar way. In the $5\msun$ case Na is mainly produced
in hot-bottom burning, by p-capture on dredged-up \nezw.

In the previous section it was argued that due to the large primary
production of CNO and many other species the $Z=0$ model predictions
would also apply approximately to stars with ultra-low metallicity as
well. Neutron capture reactions play an important role, and need to be
included in yield-predictions, together with a neutron-sink
approximation for species not explicitely included in the network.
This has been done in a qualitatively similar way by both
\citet{herwig:04a} for the $Z=10^{-4}$ models and \citet{siess:02} for
their $Z=0$ models, and the abundance evolution is quantitatively
similar. Both model sets predict for non-hot-bottom burning cases
large primary production of C and O, while the \nvi\ abundance does not
change. \nezw\ is produced in both cases in the He-shell flashes, and
in both cases the final \nezw\ exceeds that of \nvi. The evolution of
\nadr, \mgfu\ and \mgse\ is qualitatively the same too. In particular
the ratio of these three isotopes is quantitatively the same in the
last computed model of the $Z=0$ sequence and the final AGB model at
$Z=10^{-4}$. This is in particular interesting as these low-mass AGB
models predict that $\mgse:\mgfu > 1$, contrary to the signature of
hot-bottom burning.

One of the largest uncertainties is the adopted mass loss. While mass
loss in AGB stars of solar metallicity can be constraint
observationally, this has not been possible for extremely low
metallicity. Other uncertainties relate to mixing in EMP AGB stars.
\citet{herwig:03c} has shown that convection induced extra-mixing,
like exponential overshoot that is used to generate a \cdr\ pocket for
the \spr\ in higher-metallicity models, may lead to vigorous H-burning
during the third dredge-up.  Depending on the efficiency of such
mixing the third dredge-up may turn into a flame-like burning front,
leading to very deep core penetration. This may significantly impact
the formation and effectiveness of a \cdr\ pocket for the \sprn\ 
\citep{goriely:04}.

\section{Near-field cosmology application of stellar yield calculations}
\label{sec:nfc}

Accurate stellar yields for intermediate mass stars are requested by
another emerging field, near-field cosmology
\citep{bland-hawthorn:00,freeman:02}. The baryon (stellar) halo of the
Milky Way retains a fossil imprint of the merging history of the
galaxy. Different merging components, like the infalling dwarf
galaxies with a range of masses, have different star formation
histories that translate into different abundance signatures of the
member stars of these components. The current surge of spectroscopic
multi-object capabilities at large telescopes will likely
significantly enhance the importance of this approach.

An example is the recent work by \citet{venn:04} in which abundances
of halo stars are compared with abundances of stars in satellites,
dwarf galaxies trapped in the potential well of the Galaxy. Here, the
basic idea is that dwarfs are merging with a galaxy at different times
at which point star formation and chemical evolution stops. Satellites
that survive until the present day should show the signature of more
evolved chemical evolution, for example including the \spr\ elements
associated with the long-lived low-mass stars. In contrast, halo stars
that are the dispersed members of dwarf galaxies that have merged at
an earlier time in the evolution of the galaxy, have less evolved
chemical evolution patterns, for example showing more clearly the
patterns of nucleosynthesis in massive star evolution.

Potentially, the implications that can be derived from comparison of
halo stars and satellite members may include some fundamental
questions of nucleosynthesis itself. The data presented in
\citet{venn:04} show that $\mem{[\alpha/Fe]}$ is systematically
smaller in stars belonging to satellites compared to their halo
counterparts. This can be explained in the $\Lambda$ cold dark matter
model that predicts that most halo stars have formed in rather massive
dwarf galaxies, which merged with the galaxy a long time ago
\citep{robertson:05}. These stars show the signature of truncated
chemical evolution, dominated by the yields of supernovae type II and
their significant $\alpha$-element contribution and moderate Fe
ejecta. Stars in present, less massive satellite galaxies show the
signature of chemical evolution components that take more time, like
the SN Ia. These events add Fe but little $\alpha$-elements, and their
$\mem{[\alpha/Fe]}$ ratio is therefore smaller.

Abundances of other elements could be compared too. For example,
\citet{venn:04} compare among others the abundances of Y, Ba and Eu in
halo stars and satellite dwarf galaxies. Ba is typically considered an
main-component \spr\ element (at least at $\mem{[Fe/H]} > -2$) because
the elemental solar abundance has a $81\%$ \spr\ contribution
\citep{arlandini:99}. Only the smaller remaining fraction is made in
the r-process. As discussed above, the nuclear production site of the
main-component of the s-process are low- and intermediate mass AGB
stars, in the initial mass range $1.5 < M_\mem{ini}/\msun < 3.0$. This
component of galaxy chemical evolution needs even more time to
contribute than the SN Ia. And indeed, [Ba/Fe] behaves differently
from $\mem{[\alpha/Fe]}$ in halo stars and satellite stars. [Ba/Fe] is
on average higher in the satellite stars than in halo stars, with a
considerable spread.  This is consistent with the framework of
truncated chemical evolution of systems that were early disrupted in
merger events, and with the understanding of stellar evolution and
nucleosynthesis that $\alpha$-elements are predominanty made in
short-lived massive stars, while Ba originates in rather old
populations. It is then extremely interesting to consider the
r-process element Eu, which in the solar abundance distribution has a
very small \spr\ contribution of only $5.8\%$. This element does not
behave like $\alpha$-elements. Instead, [Eu/Fe] is on average the same
in halo stars and dwarf galaxy stars, however with a larger spread in
the latter. In order to remain in the proposed scenario of why
abundances in the halo differ from those in the nearby dwarf galaxies
one would then have to assume that at least an important fraction of
the r-process elements does not originate in SN II, which eject their
yields on a short time scale. Instead, one would have to assume that
Eu in the satellite stars comes from a source that releases the ejecta
on a time scale comparable to or longer than the Fe production in SN
Ia. Such sources could be the accretion induced collapse
\citep{fryer:99} or the collapse of a super-AGB star in the initial
mass range $8 \dots 10\msun$, if the core mass grows to the
Chandrasekhar limit \citep{nomoto:84}. It would certainly go to far at
this point to draw any further conclusion, for example in conjunction
with the two r-porcess source model proposed by \citet{qian:03}. 

It is clear, that the full potential of using the abundances patterns
of stars to reconstruct the formation of galaxies in their
cosmological context can only be reached with detailed yield
predictions including all masses and reaching down to extremely low
metallicity. Y is another example that reinforces this notion. In the
solar abundance distribution it has a $92\%$ fraction from the
main-component \spr\ originating in low-mass stars.  However, it is
also the termination point of the weak \spr\ from massive stars, and
it does have some r-porcess contribution. In the data presented by
\citet{venn:04} it behaves -- with regard to the difference between
halo and satellite stars -- like the $\alpha$-elements, which in this
context would imply an important contribution from massive stars.

\section{The origin of nitrogen in the early universe}
\label{sec:nitrogen}
 \begin{figure}[t]
\epsfxsize=11cm
\center{ \epsfbox{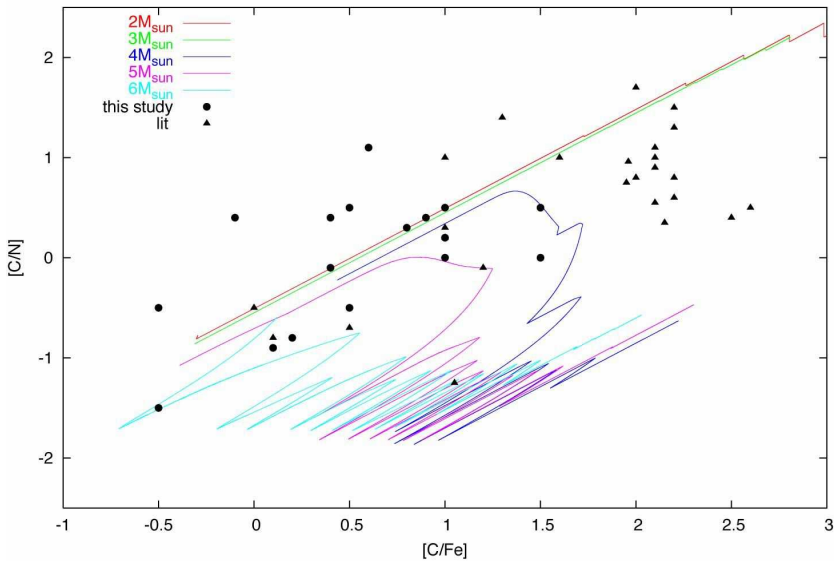}}
\caption{ \label{fig:nitrogen} 
  Observed nitrogen and carbon abundances \citep[including literature
  data from][]{barbuy:97,preston:01,aoki:01,lucatello:02} compared to
  the surface abundance evolution of AGB model tracks. }
\end{figure}
Nitrogen is the fifth most common element in the universe. In stellar
evolution of very low metallicity the primary production is of
particular importance.  Strictly from a nucleosynthesis point of view
nitrogen is always made in a well established sequence of events,
involving first the triple-$\alpha$ process that makes \czw\ and then
two subsequent proton captures that are part of the CN-cylce. In a
stellar model sequence the problem is to identify the mixing processes
that can account for this sequence of events. How can \czw, that is
made in the He-burning that requires higher temperatures, be brought
into layers of lower temperature where H-burning still takes place.
Or, how can protons be mixed down into the He-burning layers and be
captured by \czw\ to eventually form \nvi, but then not be exposed to
$\alpha$-captures that would convert \nvi\ into \nezw. Because the
primary production of nitrogen depends so much on stellar mixing, this
element assumes a key role in understanding stellar nucleosynthesis
at extremely- and ultra-low metallicity.

Nitrogen is certainly present in the most metal-poor stars in the
Milky Way \citep{laird:85}. According to standard stellar evolution
and nucleosynthesis models primary N comes from intermediate mass (IM)
AGB stars \citep{herwig:04a}, with perhaps a small contribution from
rotating massive stars \citep{meynet:02}. In AGB stars the repeating
sequence of thermal pulses induce convective mixing events that lead
to the prodution of N. C is made in the He-shell flash. Dredge-up
after the flash mixes that C into the envelope, and in massive AGB
stars the C in the envelope is then transformed into N.  The low
[N/$\alpha$] abundances observed in some damped Lyman-$\alpha$ systems
\citep{pettini:99} have prompted suggestions that the IMF may be
biased in favor of massive stars in some systems \citep{prochaska:02},
or that these systems represent the earliest stages of chemical
evolution of massive stars only \citep{centurion:03}.

Observations of N and C abundances in C-rich EMP stars with
\spr-signature provide another way to study the N production at low
metallicity. About 30$\%$ of all extremely metal-poor stars ([Fe/H] <
-2) are strongly C- and to a lesser extent N-enhanced
(\kap{sec:CEMP}). In particular the overabundance of C and of N in
addition to their s-process signature has lead to the assumption that
the AGB star progenitor of the current white dwarf companion to the
CEMP-s star is responsible for the observed abundance pattern. Because
of the established binary nature and the \spr-signature the CEMP-s
stars are assumed to be poluted by the individual AGB stars that are
the progenitors of their present white dwarfs companions.

In \abb{fig:nitrogen} the [C/N] and [C/Fe] ratios of literature
data are shown, together with the AGB model prediction of $2$ -
$6\msun$ tracks. The literature data show systematically lower [C/N]
ratios than what is expected by $2$ or $3\msun$ models that dredge-up
C, but do not produce N. This poses the question of the primary origin
of the N in theses stars.

Models of initial masses between 4
and 6 feature efficient hot-bottom burning (HBB) which turns most
dredged-up primary carbon into nitrogen resulting in [C/N]<-1 and
[C/Fe]<1.5. However, none of the literature data seem to show the very
low [C/N] ratio that would be expected for an EMP star that happened
to have an AGB companion in the $4 - 6\msun$ initial mass range.  We
have carried out observations of EMP targets with 0.5<[C/Fe]<1 using the
CH and the NH band for abundance determination
\citep{johnson:05,herwig:04g}. Specifically we wanted to find out
whether the paucity of EMP stars with [C/N]<-1 is a selection effect
or a systematic observational bias imposed by the abundance indicators
employed in previous studies.  While C-strong stars can be identified
by the G-Band at $4305$\AA, the strong NH band at $3360$\AA is never
observed in the medium-resolution surveys that provide the targets for
detailed abundance studies \citep[e.g.\ ][]{beers:92}. Only the CN bands
at $3883$\AA and $4215$\AA are included, and CN lines are not strong
in N-rich stars unless C is also enhanced by a large amount.
Therefore, studies of C and N abundances have favored C-rich stars
that are easily identified, and may have missed stars that have
$\mem{[C/Fe]}\sim 0 - 1$, but are more rich in N. Thus the
literature data in \abb{fig:nitrogen}, drawn from high-resolution
follow-up of the medium-resolution candidates, may be the result of a
strong observational bias against finding N-rich stars.

In order to overcome this bias we obtained spectra for
18 new stars.  In \abb{fig:nitrogen}, we plot the preliminary
estimates of the [C/N] ratios of 18 stars, based on observations of
the NH bands during 6 nights at CTIO/KPNO. We did not find any stars
with low [C/N], and the analysis of additional data is underway to put
our findings on a more robust statistical basis. This leaves us with
two important open questions: (1) Where are the EMP stars polluted by
massive AGB stars? and (2) Where does the N in the CEMP stars come
from that we do observe? While we do not have any idea at this point
of what the answer to the first question could be, there are a number
of possibilities to address the absence of primary N. These include
rotationally induced mixing before the AGB phase (\kap{sec:z00}), the
H-ingestion flash (\kap{sec:hflash}), and extra-mixing in AGB stars
\citep{nollett:03} as envoked in models of RGB stars to account for
abundance anomalies in globular cluster member stars
\citep{denissenkov:03b}. In any of these cases the interpretation of N
abundances in metal-poor systems of all kinds may have to be re-evaluated.

\section{H-ingestion flash and born-again evolution}
\label{sec:hingba}
 \begin{figure}[t]
\epsfxsize=11cm
\center{ \epsfbox{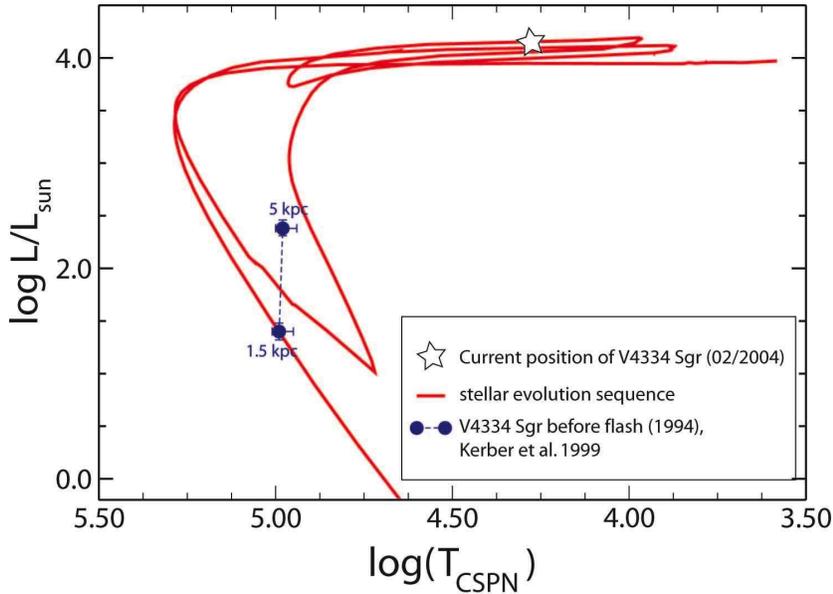}}
\caption{ \label{fig:HRDsak} Observational position of Sakurai's
  object (V4334 Sgr) at different times, and evolutionary sequence of
  a 0.604\msun\ central star of a planetary nebula track
  \citep{hajduk:05} with a very late thermal pulse (H-ingestion flash)
  with modified convective mixing velocities according to
  \citet{herwig:01a}.}
\end{figure}
Nuclear production in the AGB stellar interior can become observable
at a later time in the evolution, during the post-AGB phase, on the
surface of those central stars of planetary nebulae that become
H-deficient. Here H-deficient means in most cases that the mass
fraction of H as observed in the stellar spectrum is less than
$\approx 2\%$. This leads to additional important constraints on the
evolution of AGB stars, which is in particular important to study AGB
evolution at extremely low metallicitiy.

H-deficient central stars of planetary nebulae (CSPN) as well as very young
white dwarfs have been analysed in detail
\citep[e.g.][]{koesterke:01,werner:01,hamann:03} and new important
information about the abundances of these stars is yet to be revealed.
Stellar evolution predicts that about $25\%$ of all post-AGB stars
will eventually loose all their small remaining H-rich envelope mass
of the order $10^{-4} \msun$ and expose the bare H-free cores. The
carbon, oxygen and helium rich nature of these objects is evident from
their Wolf-Rayet or PG\,1159 type emission line spectra, and reflects
nuclear processed material that has been built up during the stars
progenitor evolution. Although there is considerable spread in the
observed abundances of these stars, the pattern can be summarized in
mass fraction as He $\sim$ C $\sim 0.4$ and $\mem{O}=0.08 \dots 0.15$.
 
 These H-deficient CSPN are important for AGB evolution modeling,
 because they provide a unique opportunity to study directly the
 nuclear processing shells in AGB stars. In order to explain the
 evolutionary origin \citet{iben:83a} introduced the born-again
 evolution scenario . The star evolves off the AGB, and becomes a
 hydrogen-rich central star and eventually a very hot, young white
 dwarf. However, the He-shell may still be capable to ignite a late
 He-shell flash, and in that case the star retraces its evolution in
 the HRD, back into the giant region (\abb{fig:HRDsak}). As a result
 of hydrogen-ingestion into the He-convection and rapid burning, or by
 mixing from the emerging convective envelope (or because of both) the
 surface abundance of such born-again stars will be extremely
 H-deficient or even H-free.
 
 Stellar models of this evolutionary origin scenario connect the
 surface abundance of the Wolf-Rayet central stars and the PG1159
 stars with the intershell abundance of the progenitor AGB star. The
 intershell layer between the He- and the H-shell is well mixed during
 each He-shell flash. This zone contains the main nuclear production
 site of the progenitor AGB star, and the abundance of this zone
 reflects contributions from both He- and H-shell burning
 (\abb{fig:kipp}). As a result of the born-again evolution these
 layers become visible at the surface of the resulting H-deficient
 CSPN.

The initial models of the H-deficient central stars
 by Iben and collaborators showed qualitatively that the born-again
 scenario could account for high He and C abundances, as these
 elements were abundant in the intershell of the AGB progenitor model
 they used. However, they could not account for the high observed
 oxygen abundance. \citet{herwig:97} were the first to propose that
 the solution could be non-standard mixing during the AGB
 evolution. Models with overshooting at the bottom of the pulse-driven
 convection zone do not only feature higher temperatures at the bottom
 of that layer (\kap{sec:spr-mix}), but they also show higher C and in
 particular O abundances in the PDCZ. Subsequently, \citet{herwig:99a} has
 explored in detail how the various abundances depend on the overshoot
 efficiency and other details. In essence, overshooting brings AGB
 intershell abundances of He, C and O in very good quantitative
 agreement with the observed abundances of Wolf-Rayet type central
 stars and PG1159 stars, in the framework of the born-again
 evolution. More effort is needed to consolidate these constraints on
 the intershell abundance with the possibly tight upper limits on
 overshooting at the bottom of the PDCZ that the \sprn\ may provide
 (\kap{sec:spr-mix}).
 
 The connection of the surface abundances of the hot PG1159 stars and
 the progenitor AGB intershell abundance has been reinforced recently
 by several new observational findings. These include a substantial
 overabundance of Ne \citep{werner:04}, F abundances ranging from
 solar up to 250 times solar \citep{werner:05b} in good agreement with
 AGB nucleosynthesis predictions by \citet{lugaro:04a}, and
 Fe-deficiencies of at least 1 dex compared to solar \citep{miksa:02},
 which may reflect the depeletion of Fe in the AGB intershell due to
 \spr\ n-captures. In that case the Fe/Ni ratio should be low, as in
 fact observed in Sakurai's object. This born-again star is the
 smoking gun of the very late thermal pulse scenario and links this
 evolutionary scenario with the H-deficient PG 1159 and Wolf-Rayet
 type central stars \citep{duerbeck:00,herwig:01a}. Much of the
 observed abundance pattern of Sakurai's object as well as the rapid
 evolutionary time scale has been reproduced quantitatively with
 stellar models \citep{asplund:99a,herwig:01a}. Similar to the PG 1159
 stars, Sakurai's object shows Fe depletion of about 1dex compared to
 solar.  The low ratio of $\mem{(Fe/Ni)} = 3$, shows that Sakurai's
 object shows surface material that has been directly irradiated with
 neutrons \citep{herwigIAU209P4_01}. New radio observations show that
 Sakurai's object has started reheating again, on its second evolution
 into the CSPN \citep{hajduk:05}. This marks a new phase of the
 evolution of this star, and confirms the concept of convective mixing
 efficiency modified by nuclear burning \citep{herwig:01a}.
  
\section{\sprn\ and AGB stars}
\label{sec:spr}
 \begin{figure}[t]
\epsfxsize=11cm
\center{ \epsfbox{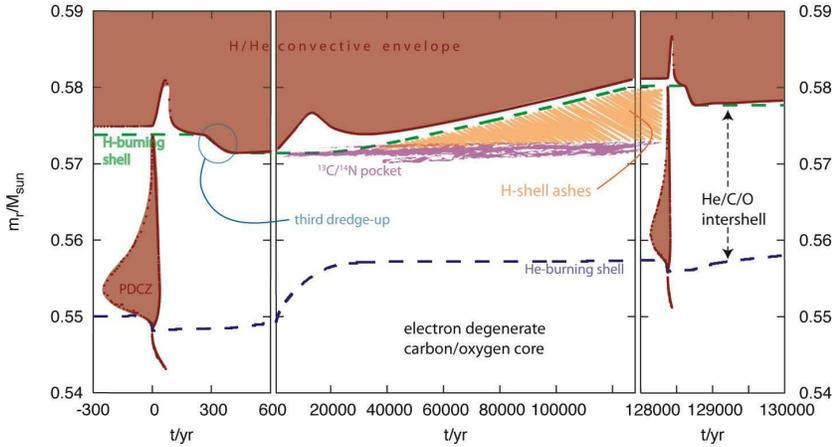}}
\caption{ \label{fig:kipp} 
  Evolution of convection zones (red regions) and burning shells
  (dashed lines) in 1D stellar track including two He-shell flashes
  and the quiescent H-burning phase inbetween. The purple indicates
  the location of the \cdr-rich region that contains the bulk
  \spr-production. The orange region contains the ashes of the H-shell
  that are engulfed by the following He-shell flash convection zone.}
\end{figure}

The \sprn\ is the origin of half of all elements heavier than iron
\citep{arlandini:99}. It is also important for isotopic ratios and in
some cases elemental abundances of lighter elements - in particular at
extremely low metallicity. The heavy elements are made by the \sprn\
through neutron captures that are slow compared to the competing
$\beta$-decay. Starting from the abundant iron group elements it
follows closely the valley of stability in the chart of isotopes. It
is characterized by neutron densities $N_\mem{n}<10^{10}
\mem{cm^{-3}}$. Observations and theory agree that the nuclear
production site of the main and strong component of the \sprn\
($90<A<204$) originates in low mass Asymptotic Giant Branch (AGB)
stars. The weak component below the first \spr\ peak at $A=90$ is
produced during He and C burning in massive stars. Here we deal only
with the \spr\ in AGB stars.  

The \spr\ is important in the context of the second stars for two
reasons. First, the observed elemental abundance distribution of the
trans-iron elements is at any given metallicitiy a mix of s- and
r-process contribution. In order to disentangle the two contributions
in very-low metallicity stars, the \spr\ in this metallicity regime
needs to be understood. This may than enable progress in identifying
the conditions for the r-process. Second, \spr-branchings can be used
to probe the physics for stellar mixing (\kap{sec:spr-diag}), which is
particularly usefull when modeling the evolution of EMP stars.

The \sprn\ in low-mass AGB stars has two neutron sources. The main
source is \cdr, that forms via the $\czw(\p,\gamma)\ndr(\beta^+)\cdr$
reaction. At the end of the dredge-up phase (\abb{fig:kipp}) after
the He-shell flash, when the bottom of the H-rich convective envelope
has penetrated into the \czw\ rich intershell layer, partial mixing at
this interface would create a thin layer providing simultaneously the
required protons and \czw. The \sprn\ in the \cdr\ pocket is
characterized by low neutron densities ($\log N_\mem{n} \sim 7$) that
last for several thousand years under radiative, convectively stable
conditions during the quiescent interpulse phase. The physics of
mixing at the H/\czw\ interface at the end of the third dredge-up
phase has not yet been clearly identified (see below). Most likely it
is some type of convection induced mixing beyond the convection
boundary.

The $\nezw(\alpha,\n)\mgfu$ reaction requires the high temperatures
that can be found at the bottom of the pulse-driven convection zone
(PDCZ) during the He-shell flash ($T>2.5\cdot10^8\kelv$). The neutrons
are released with high density ($\log N_\mem{n} \sim 9 \dots 11$) in a
short burst \citep{gallino:97b}. These peak neutron densities are
realised for only about a year, followed by a neutron density tail
that lasts a few years, depending on the stellar model assumptions.
The current quantitative modeling of the \spr\ uses the thermodynamic
output from a stellar evolution calculation including mass loss as
input for nucleosynthesis calculations with a complete \spr\ network
\citet{busso:99}. The post-processing accounts for both the \cdr\ 
neutron source as well as for the \nezw\ source, and mixes the
different contributions according to the information provided by the
stellar evolution calculations. The free parameter of the model is the
\cdr\ abundance in the \cdr\ pocket that is proportional to the
neutron exposure that results from burning the \cdr\ in the
$(\alpha,n)$ reaction. Physical mixing processes which are responsible
for bringing protons down from the envelope into the \czw-rich core to
enable \cdr\ formation, are not explicitely included in this model.

Observationally, models have to account for observed spread in
observables that are related to the neutron exposure
\citep[e.g.][]{vanwinckel:00,vaneck:03,nicolussi:98b}.  This spread is
not only evident from stellar spectroscopy, but also from SiC grain
data, that indicate that a spread by a factor of five is necessary for
the neutron exposure for a given mass and metallicity
\citep{lugaro:02b}. Currently, this spread is accounted for by a range of
different cases in each of which the \cdr\ abundance in the pocket is
assumed to be different \citep{busso:01a}. However, the physics of the
mixing that is associated with the range of neutron exposures has yet
to be identified.
  
\subsection{Convection and rotationally induced mixing for the \sprn}
\label{sec:spr-mix}
It is useful to distinguish \spr-mixing for solar and moderately
metal-poor stellar evolution, and the extremely and ultra metal-poor
cases. In the first, the assumption that \spr-mixing does not feedback
strongly into the thermodynamic evolution is generally valid, in the
latter it is generally not. Many more observational constraints exist
for the solar and moderately metal-poor \sprn, in particular the
isotopic information from the pre-solar SiC grains.

Mixing for the for the solar and moderately metal-poor \sprn\ has to
satisfy two general constraints: (1) How is the partial mixing zone of
H and \czw\ generated that eventually forms the neutron source species
\cdr, and (2) what is the origin of the observed spread in neutron
exposures.  Mixing for the \cdr\ pocket is probably related to the
penetrative evolution of the bottom of the convective envelope during
the third dredge-up.  Possible mechanisms include exponential
diffusive overshooting \citep{herwig:97,herwig:99a}, mixing induced by
rotation \citep{langer:99}, and mixing by internal gravity waves
\citep{denissenkov:02}. Each of these effectively leads to a
continuously and quickly decreasing mixing efficiency from the H-rich
convection zone into the radiative \czw-rich layer, and each of these
will lead to the formation of two pockets which are overlapping
\citep{lugaro:02a}.  At low H/\czw\ ratios H-burning is
proton-limited, and protons will make \cdr\ via the $\czw(\p,\gamma)$
reaction but no (or little) \nvi. At larger H/\czw\ ratio a \nvi\
pocket forms.  The maximum abundance of both \cdr\ and \nvi\ depends
on the \czw\ abundance in the intershell. Therefore, the conditions in
the \cdr\ pocket are not independent of, for example, the mixing at the bottom of
the PDCZ. \citet{lugaro:02a} derive the relationship
$\tau_\mem{max}=1.2 X(\mem{\czw})_\mem{IS}+0.4$, where
$X(\mem{\czw})_\mem{IS}$ is the intershell \czw\ mass fraction and
$\tau_\mem{max}$ is the maximum neutron exposure reached in the \cdr\
pocket. The observations of H-deficient post-AGB stars described in
\kap{sec:hingba} require that the \czw\ abundance in the PDCZ of AGB
stars is about $40\%$. This value is reproduced in AGB stellar
evolution models with overshooting at the bottom of the PDCZ, and
about twice as large as in models without overshooting.

Apart from the maximum amount of \cdr\ in the pocket the mixing
process must produce a partial mixing layer of the right mass $\Delta
M_\mem{spr}$ which eventually contains \spr\ enriched material.  In
the exponential overshooting model of \citet{herwig:97} the mixing
coefficient is written as $ D_{\rm OV} = D_0 \exp{\left( \frac{-2
      z}{f_\mem{ov} \cdot H_{\rm p}}\right)}$, where $D_0$ is the
mixing-length theory mixing coefficient at the base of the convection
zone, $z$ is the geometric distance to the convective boundary,
$H_{\rm p}$ is the pressure scale height at the convective boundary,
and $f_\mem{ov}$ is the overshooting parameter. If applied to core
convection $f_\mem{ov}=0.016$ reproduces the observed width of the
main sequence. \citet{lugaro:02a} finds that $f_\mem{ov}=0.128$ at the
bottom of the convective envelope generates a large enough \cdr\ 
pocket. However, the maximum neutron exposure in the \cdr\ pocket of
the overshooting model is $0.7\cdots0.8\mem{mbarn^{-1}}$, while in the
non-overshooting models this value is $0.4\mem{mbarn^{-1}}$.
Correspondingly \citet{lugaro:02a} obtained only negative values for
the logarithmic ratio [hs/ls]\footnote{hs and ls are the heavy and
  light \spr-indices which are the average of the abundances of
  elements around the neutron magic numbers 50 and 80.}, while the
overshooting model with the larger neutron exposure predicts
[hs/ls]$\sim 0$. This compares to an observed range of
$-0.6<\mem{[hs/ls]}<0.0$ for stars of solar metallicity \citep[see][for a
compilation of observational data]{busso:01a}. Models with
overshooting at all convective boundaries can reproduce only the
largest observed hs/ls ratios, indicating that the neutron exposure in
the \cdr\ pocket in these models is at the maximum of the
observationally bounded range. 

Overshooting alone is not able to account for all features of
\spr-mixing. In particular there is no mechanism to acount for the
spread in neutron exposures within the overshooting framework.
Rotation, however, may induce a range of mixing efficiencies for a
sample of stars with otherwise identical parameters. Models of
rotating AGB stars were presented by \citet{langer:99}, and the \sprn\
was analyzed in detail by \citet{herwig:02a} and \citet{siess:04}. The
implementation of rotation for the AGB models was the same as the one
that had been used previously to construct rotating models of massive
stars \citep{heger:00}. This implementation yields a \cdr\ pocket
generated by shear mixing below the envelope convection base, however
an order of magnitude smaller than what is needed in the partial
mixing zone of a non-rotating model to reproduce the observed
overproduction in stars. The second important finding is that shear
mixing which initially generates the small \cdr\ pocket, prevails
throughout the interpulse phase, even when the base of the convection
is receding in mass after H-shell burning has resumed.  When the
dredge-up ends, the low-density slowly rotating convective envelope and
the fast rotating compact radiative core are in contact and mixing is
induced through shear at this location of large differential
rotation. This radial velocity gradient remains as a source for shear
mixing at exactly the mass coordinate of the \cdr\ pocket with
important consequence for the \spr. Shear mixing during the interpulse
phase swamps the \cdr\ pocket with \nvi\ from the pocket just
above. By the time the temperature has reached about $9\cdot
10^{7}\kelv$ and \cdr\ starts to release neutrons via
$\cdr(\alpha,\n)\ose$, \nvi\ is in fact more abundant than \cdr\ in
all layers of the \cdr\ pocket. \nvi\ is a very efficient neutron
poison. It has a very large $\nvi(\n,\p)\cvi$ rate and simply steels
neutrons from the iron seed. As a result, the neutron exposure is only
$\sim 0.04\mem{mbarn^{-1}}$, about a factor of ten too small to
generate the observed \spr\ abundance distribution . The detailed
post-processing models of current rotating AGB stars showed that they
are not capable to account for the observed \spr\
overabundances. Parametric models show that weaker shear mixing during
the interpulse leads to a weaker poisoning effect. For very small
poisoning effects the neutron exposures are still large enough to
reproduce some observations.

While neither rotation or overshooting alone provide the right amount
of \spr-mixing it is interesting to consider a combination of
these. In essence the idea is that overshooting would provide mixing
for the formation for the \cdr\ pocket and for a larger \czw\
abundance in the PDCZ to obtain a large neutron exposure in the \cdr\
pocket. Then, shear mixing during the interpulse could add some poison
and result in the observed spread. Magnetic fields have not yet been
considered in AGB stellar evolution. However, one may assume that
qualitatively magnetic fields will add coupling between the fast
rotating core and the slowly rotating envelope and provide additional
angular momentum transport. Models including this effect could result
in smaller shear mixing than predict by the current rotating AGB
stellar models.

It is interesting to note that such an effect of magnetic fields may
also help to reconcile the predicted rotation rates of AGB cores of
$\sim 30 \mem{km/s}$ with the rotation rate determinations of white
dwarfs. Spectroscopic determinations of rotation rates of white dwarfs
of spectral type DA can not rule out such values, but most are also
consistent with zero or very low rotation
\citep{koester:98,heber:97}. However, asterioseismological
measurements of WD rotation rates clearly yield smaller values in the
range $0.1 < v_\mem{rot}/(\mem{km/s}) <1 $ \citep[see references
in][]{kawaler:03}.

\subsection{\sprn\ as a diagnostic tool}
\label{sec:spr-diag}
 \begin{figure}[t]
\epsfxsize=11cm
\center{ \epsfbox{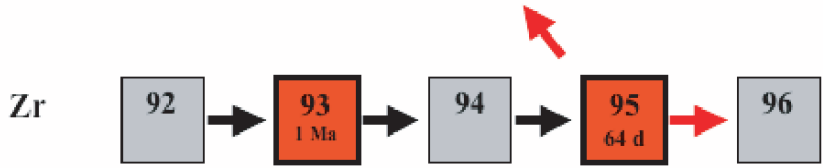}}
\caption{ \label{fig:zr}
  Small section of the chart of nuclides, showing the branching at
  \zrfu. The \sprn\ proceeds by adding neutrons to both stable (grey) and
  unstable (red) species. Radioactive species like \zrfu\ can either
  $\beta$-decay or capture a neutron.  }
\end{figure}
The high precision information on the pre-solar meteoritic SiC grains
provide isotope ratio measurements that allow to probe the conditions
at the \spr\ nuclear production site in more detail. One example is
the highly temperature dependent nucleosynthesis triggered by the
release of neutrons from \nezw\ at the bottom of the PDCZ.  The
temperature at the bottom of the PDCZ correlates with the efficiency
of extra mixing like overshooting at the bottom of this convection
zone \citep{herwig:99a}.  Neutrons in the PDCZ are generated by the
$\nezw(\alpha,\n)\mgfu$ reaction. For larger temperatures the neutron
density is higher. Isotopic ratios that enclose a branch point isotope
in the \spr\ path will be more neutron heavy for higher neutron
densities. An example is the \zrse/\zrvi-ratio, that is set by the
branching at the radioactive \zrfu\ (\abb{fig:zr}). 
 For low neutron densities \zrfu\ decays. For
$N_\mem{n} > 3\cdot10^8 \mem{cm^{-3}}$ $\zrfu(\n,\gamma)\zrse$ becomes
significant, hence \zrse\ is produced. If the temperature is larger
the neutron density is larger and the \zrse/\zrvi\ ratio, which can be
measured in SiC grains, is larger as well.

\citet{lugaro:02b} have studied the measured isotopic ratios of Mo and
Zr as well as Sr and Ba from SiC grains. They evaluate the sensitivity
of their results in terms of nuclear reaction rate uncertainties. All
branchings that are activated by the \nezw\ neutron source depend on
the still uncertain $\nezw(\alpha,\n)\mgfu$ rate. In addition, the
\zrse/\zrvi\ ratio depends on the neutron cross-section of the
unstable isotope \zrfu. Like for almost all radioactive nuclei the
$(\n,\gamma)$ rate of \zrfu\ is not measured.  The theoretical
estimates vary from Maxwellian-averaged cross sections of $20\mem{mb}$
\citep{beer:92} to $140\mem{mb}$ (JENDL-3.2).  In order to make full
use of the potent method of using the \sprn\ as a diagnostic tool it
is critical that n-cpature rates of the radioactive \spr-branch point
nuclides are measured. Different \spr-branchings are sensititv to
different mixing processes, including those that may be induced by rotation.

The \spr\ as a diagnostic tool can
provide information on mixing processes that are potentially relevant
for the evolution of stars of all masses, including the progenitors of
supernovae. In particular the details of the initial model for a supernova
calculations determines important properties of the
explosion, like asymmetries, or the final fate as black hole or
neutron star \citep{young:05}. 

\section{Carbon star formation and nuclear reaction rate input}
\label{sec:enrt}
In order to understand C-rich extremely metal-poor (CEMP) stars the
formation of C-rich AGB stars has to be understood in a quantitative
way. AGB stars become C-rich because of the third dredge-up, which
mixes C-rich material from the intershell into the envelope
(\abb{fig:kipp}). The evolution of C and O in AGB stars and the
problems related to modelling the third dredge-up are discussed in
 \citet{herwig:04c}. In summary, the third dredge-up is now
obtained in 1D stellar evolution calculations in sufficient amount and
at sufficient low core mass. These calculations take into account
convection-induced mixing into the stable layers in a time- and depth
dependent way, and use high numerical resolution. Uncertainty is
introduced by the mixing length parameter \citep{boothroyd:88}. 

Understanding the dredge-up properties of AGB stars is important,
because the dredge-up dependnent yield predictions for low and
intermediate mass stars enter models for galaxy chemical evolution.
AGB stars serve as diagnostics for extragalactic populations, and for
this purpose the conditions of the O-rich to C-rich transitions needs
to be known. C-rich giants are the brightest infrared
population in extra-galactic systems. Finally, the
envelope enrichment of AGB stars with the s-process elements is
intimately related to the dredge-up properties of the models.
 \begin{figure}[t]
\epsfxsize=11cm
\center{ \epsfbox{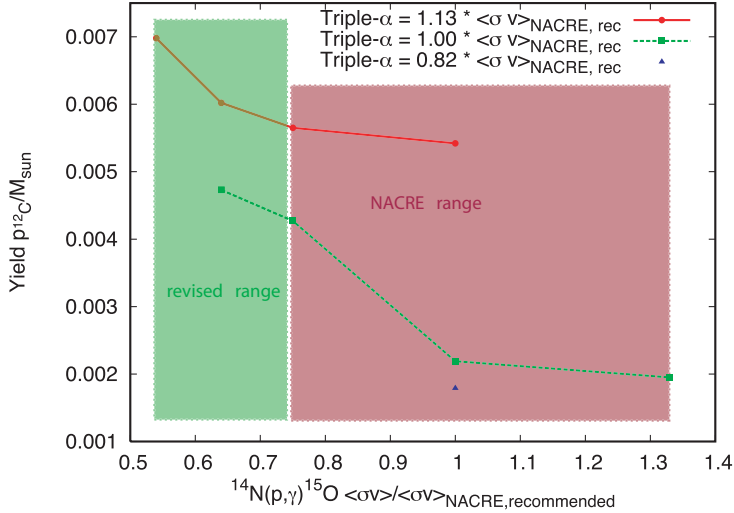}}
\caption{ \label{fig:enrt} 
  Carbon yield as a function of nuclear physics input. Each point
  gives the yield. from one full $2\msun$, $Z=0.01$ TP-AGB stellar
  evolution sequence with the indicated choice for the
  $\nvi(\p,\gamma)\ovi$ and the triple-$\alpha$ reaction. The yield of
  a species is the its abundance in the ejected material minus the
  initial abundance integrated over the mass loss. }
\end{figure}

Recently, it has been shown that uncertainties in nuclear reaction
rates propagate in a significant way into the dredge-up and thereby
yield predictions. \citet{herwig:04b} calculated an extensive grid of
$2\msun$, $Z=0.01$ tracks for combinations of rates for the
$\nvi(\p,\gamma)\ovi$, the triple-$\alpha$ and the
$\czw(\alpha,\gamma)\ose$ rate within the errors given in the NACRE
compilation \citep{angulo:99}. The main result was that dredge-up and
C yields are larger for lower $\nvi(\p,\gamma)$ rate and for larger
triple-$\alpha$ rate. The $\czw(\alpha,\gamma)$ rate plays a less
important role. It was also found that nuclear physics work since the
1999 NACRE compilation require a downward revision of the
$\nvi(p,\gamma)$ rate, by almost a factor of 2. \abb{fig:enrt} shows
the revised range for this rate and the yields from the 
calculations assuming the uncertainty range of the nuclear reaction rates. 

\section{Conclusions}
\label{sec:concl}
Extremely metal-poor stars are an emerging field of astrophysical and
astronomical research, pushing the limits of observations, and theory
and numerical simulation. Potentially, much can be learned about
challenging questions of astrophysics. How do galaxies like our Milky
Way form? How did the first stars and their cosmological environment
form and evolve? Some exciting clues about the evolution of stars come
from the smallest astrophysical bodies, pre-solar stardust extracted
from primitive meteorites. Thus stellar evolution and nucleosynthesis
connects nearby phenomena of planetary system formation with star
formation and evolution in the earliest time of the universe as
recorded in the element distribution patterns of the most metal-poor
stars.

In the future the quantity and quality of spectroscopic data of stars,
in particular the valuable most metal-poor stars, will increase
dramatically, due to multi-object spectroscopy and large spectroscopic
surveys. It is already becoming clear that this data can be put to
full use only with qualitatively and quantitatively improved
simulations of nuclear production in stars, including low and
intermediate mass and massive stars.

\subsection*{Acknowledgements} 
I would like to thank D.\ Schoenberner for nominating me for the
Ludwig-Biermann award, and for directing my interest to stellar
evolution and nucleosynthesis as my \emph{Doktorvater}. I am very
grateful to N.\ Langer and D.\ VandenBerg for their continuing
support. I would also like to thank my colleagues at the Theoretical
Astrophysics group (T-6) at Los Alamos National Laboratory, in
particular C.\ Fryer, A.\ Heger, and F.\ Timmes, as well as R.\
Reifarth at LANSCE-3, who contribute to a very stimulating
atmosphere. Finally, I would like to thank A.\ Font and B.\ O'Shea for
stimulating discussions, that have contributed to some views expressed
in this article.  This work was funded under the auspices of the U.S.\
Dept.\ of Energy, and supported by its contract W-7405-ENG-36 to Los
Alamos National Laboratory.


\end{document}